

\input harvmac.tex
\overfullrule=0pt
%
%
\Title{\vbox{\baselineskip12pt
\hbox{IASSNS-HEP-92/66}
\hbox{LAVAL PHY-26-92}
\hbox{hepth@xxx/9210080}}}
{\vbox{
\centerline{ On The Hamiltonian Structures}
\vskip5pt
\centerline{and}
\vskip5pt
\centerline{The Reductions of The KP Hierarchy$^{\scriptstyle *}$ }}}
\footnote{}{Talk given by D.A.D. at the NSERC-CAP Workshop on \it Quantum
Groups, Integrable Models and Statistical Systems, \rm Kingston, Canada
July 13-17 1992.}
\vskip.5cm
\centerline{Didier A.~Depireux$^{\dag}$ and Jeremy Schiff$^{\ddag}$}
\vskip.5cm
\centerline{${}^{\dag}$          D\'epartement de Physique,}
\centerline{\phantom{${}^{\dag}$}Universit\'e Laval,}
\centerline{\phantom{${}^{\dag}$}Qu\'ebec, Canada G1K 7P4.}
\vskip.5cm
\centerline{${}^{\ddag}$         School of Natural Sciences,}
\centerline{\phantom{${}^{\dag}$}Institute for Advanced Study,}
\centerline{\phantom{${}^{\dag}$}Olden Lane, Princeton, NJ08540.}
\vskip .3in
{\bf Abstract:} Recent work on a free field realization of the
Hamiltonian structures of the classical KP hierarchy and of its flows is
reviewed. It is shown that it corresponds to a reduction of KP to the NLS
system. \vskip .3in

\Date{ Sep. 1992.}
%
%
\baselineskip=14pt plus 2pt minus 2pt
\def\frac#1#2{{#1 \over #2}}
\def\ha{{1 \over 2}}
\def\a{\alpha}
\def\b{\beta}
\def\d{\delta}
\def\pa{\partial}
\def\j{\jmath\,}
\def\bj{\bar\jmath\,}
\def\CH{{\cal H}}
%
\lref\KP{D.A.Depireux, Mod. Phys. Lett.~\bf A7 \rm (1992) 1825.}
\lref\Wata{Y. Watanabe, Lett. Math. Phys. \bf 7 \rm (1983) 99; Ann. Mat.
Pura Appl. \bf 137 \rm (1984) 77.}
\lref\Dick{L.A. Dickey, {\it Soliton equations and Hamiltonian systems}
  \rm, World Scientific, 1991.}
\lref\WYnine{F. Yu and Y.-S. Wu, Nucl.Phys. \bf B373\rm (1992) 713; J.M.
Figueroa-O'Farril, J. Mas  and E. Ramos, Phys. Lett. \bf B266 \rm (1991) 298.}
\lref\PRS{C.N. Pope, L.J. Romans and
X.Shen, Phys.Lett. \bf B236 \rm (1990) 173; Phys.Lett. \bf B242 \rm (1990)
401; Phys.Lett. \bf B245 \rm (1990) 72.}
\lref\Zamo{A.B. Zamolodchikov, Theor. Math. Phys. \bf65 \rm(1985) 1205;
 Adv.Stud.in Pure Math. {\bf 19} (1989) 641; A.B.
Zamolodchikov and V.A. Fateev, Nucl. Phys. \bf B280 \rm [FS18] (1987) 644;
V.A. Fateev and S.L. Lykyanov, Int. J. Mod. Phys. \bf A3 \rm(1988) 507.}
\lref\YannisII{I. Bakas and
E. Kiritsis, Nucl. Phys. \bf B343 \rm (1990) 185; Erratum \it ibid. \bf B350
\rm (1991) 512.}
\lref\WYnineteen{F. Yu and Y.-S. Wu, Phys.Rev.Lett. \bf68 \rm (1992) 2996. }
\lref\Sato{E. Date, M. Jimbo, M.
Kashiwara and T. Miwa, ``\it Proc. of RIMS Symposium on non-linear integrable
systems\rm'', eds. M.Jimbo and T.Miwa,  (World Scientific, 1983); G.Segal and
G.Wilson, Publ. IHES \bf 61 \rm (1985) 1.}
\lref\FFMR{J.M. Figueroa-O'Farril, J. Mas and E. Ramos, ``\it On the
two-boson picture of the KP hierarchy\rm'', hep-th/9207013, Bonn-HE-92-17.}
\lref\Fordy{See for instance ``\it Soliton Theory: a survey of results\rm'',
ed. A.P.Fordy (Manchester University Press 1990), p.11.}
\lref\Walter{W.Oevel, ``\it Darboux Theorems and Wronskian Formulas for
Integrable Systems I: Constrained KP flows\rm'', Loughborough University
Mathematics Report A172.}
\lref\OS{W.Oevel and W.Strampp, ``\it Constrained KP hierarchy and
Bi-Hamiltonian Structures\rm'', Loughborough University
Mathematics Report A168.}
\lref\MSBDNS{L.J. Mason and G.A.J. Sparling, Phys. Lett. \bf A137 \rm (1989)
29; I. Bakas and D.A. Depireux, Mod. Phys. Lett. \bf A6 \rm(1991) 399; J.
Schiff, PhD Thesis, Columbia University
(1991); D.A.Depireux, PhD thesis, University
of Maryland (1991).}
\lref\BX{L.Bonora and C.S.Xiong,
Phys.Lett.{\bf B285} (1992) 191; ``\it Matrix models without scaling
limit\rm'',
SISSA--161/92/EP; C.S.Xiong, contribution to this workshop.}
\lref\JerNLS{J.Schiff, ``\it The Nonlinear Schr\"odinger Equation and
Conserved Quantities in the Deformed Parafermion and SL(2,R)/U(1)
Coset Models\rm'', hep-th/9210029, IASSNS-HEP-92/57.}
\lref\Olver{P.J.Olver, {\it Applications of Lie groups to differential
equations}, Springer-Verlag, 1986.}
\lref\YWY{F. Yu and Y.-S. Wu, Phys.Lett. \bf B263 \rm (1991) 220; K.
Yamagishi, Phys.Lett. \bf B259 \rm (1991) 436.}
\lref\Chud{G.V.Chudnosky, Proc. of the Lecce Workshop 1979, Lect. notes in
Phys. \bf120\rm (1980) 104; see also Y. Cheng and Y.S.Li, Phys. Lett \bf
A157\rm (1991) 22.}

%
%

\secno=-1
\newsec{Introduction}

It is well-known that it is possible to reduce the KP hierarchy to any of the
SL(N) KdV hierarchies. The KP hierarchy has a bi-Hamiltonian structure \Dick\
and its second Hamiltonian structure has been shown \WYnine\ to be related
to $\hat W_\infty$, a centerless, non-linear deformation of the $W_\infty$
algebra of Pope, Romans and Shen \PRS. This naturally leads to the conjecture
that the $W_N$ algebras of Zamolodchikov
\Zamo, which arise as the second Hamiltonian structure of the SL(N) KdV
hierarchies, can be obtained via some sort of reduction of $\hat W_\infty$
\WYnine. While the evidence for this conjecture is convincing, an explicit
proof is still lacking.

On the other hand, inspired by the fact that the \it linear \rm algebra
$W_\infty$ possesses a representation in terms of two bosons \YannisII, Yu
and Wu \WYnineteen\ presented a two boson representation of $\hat W_\infty$.
In \KP, it was shown that this representation is related to the
reduction of the KP hierarchy to a $1+1$ dimensional integrable hierarchy.
In this talk, we recall some of the results of  \WYnineteen\ and of \KP, and
show further that the integrable hierarchy of \KP\ is related to the
well-known nonlinear Schr\"odinger (NLS) hierarchy. It seems that the fact
that KP admits such a reduction has been known  to specialists in
integrable systems for some time \OS\Chud.

Related results, albeit from a very different approach, were also
presented by C.S.Xiong at this workshop \BX.

We will be using differential $\pa \equiv \pa_x$ and
pseudo-differential $\pa^{-1}$ operators, where $\pa^{-1}$ is an integration
symbol satisfying $\pa \pa^{-1} a = \pa^{-1} \pa a = a.$ We have $\del^{-1} a
= a \del^{-1} - a' \del^{-2} + a'' \del^{-3} - \ldots$ To shorten the
expressions of the Poisson brackets, we write $\{a,b\}=c \d'$ for
$\{a(x),b(x')\} = c(x) \pa_x \d(x,x').$

\newsec{The two boson representation of $\hat W_\infty$}

\noindent{\it The KP hierarchy}

\noindent
The KP hierarchy is usually introduced following the approach of
the Japanese school \Sato;
specifically, the $r^{th}$ flow of the KP hierarchy is given by the
Lax equation
\eqn\la{{\pa\over\pa t_r} L = [ (L^r)_+,L]\qquad , \qquad\ r \in N}
where $L$ is the pseudo-differential operator
\eqn\lb{
L = \pa + u_0 \pa^{-1}+ u_1 \pa^{-2} + \ldots = \pa + \sum_{i=0}^\infty u_i
\pa^{-i-1}\quad,\quad\pa = \pa/\pa x~. }
and $(L^r)_+$ is the differential part of the $r^{th}$ power of the KP
operator $L.$
The fields $u_i$ are implicitly understood to depend on $x$ and on an
infinite number of time coordinates $t_i.$ For instance, for $r=2$ we get
$(L^2)_+ = \del^2 + 2 u_0$ and
\eqnn\lc
$$
\eqalignno{
r=2:\quad u_{0,t_2} & = 2 u_{1,x} + u_{0,xx} \quad,\cr
u_{1,t_2} & = 2 u_{2,x} + u_{1,xx} + 2 u_0 u_{0,x}\quad, \cr
u_{2,t_2} & = 2 u_{3,x} + u_{2,xx} + 4 u_{0,x} u_1 - 2 u_0 u_{0,xx}\quad, \cr
{} & \vdots  &\lc  \cr}$$
The flows defined by \la\ commute ($\del_{t_i} \del_{t_j} u_k =
\del_{t_j} \del_{t_i} u_k $) and furthermore they
are {\it bi-Hamiltonian}, i.e. \la\ can also be written as
\eqn\ld{
\del_{t_r} u_i = \{ u_i , \int \CH_{r+1} \}_1
               = \{ u_i , \int \CH_r     \}_2      }
where $\{.,.\}_1$ and  $\{.,.\}_2$ are two
Poisson brackets for the fields $u_i$ and the $\int
\CH_r$'s are some Hamiltonians. The brackets $\{.,.\}_1$
(historically the first set to be discovered
\Wata) were shown in \YWY\ to
correspond to the \it linear \rm conformal algebra $W_{1+\infty}$
with $c=0.$ The second Hamiltonian structure \Dick\ corresponds
to $\hat W_\infty$ \WYnineteen.
The Hamiltonians are given explicitly by
\eqn\lef{
\CH_r = \frac1r \ res \ L^r\quad,}
where $res \  L^r$ denotes the coefficient of $\del^{-1}$ in $L^r.$

We give the first commutators of these algebras,
as we will need their expressions shortly. The relation between the $u$
and $W$ fields will be given later. For $W_{1+\infty}$ we have
\eqnn\le
$$\eqalignno{
\{W_1,W_1\}_1 & = 0 \quad,\quad \{W_2,W_1\}_1 = W_1 \d' \quad,\cr
\{W_2,W_2\}_1 & = 2 W_2 \d' + W_2' \d \quad,\quad \{W_1,W_3\}_1 = 2 W_2
\d' + 2 W_2' \d \quad,& \le \cr }
$$
and for
$\hat W_\infty$
we have
\eqnn\lee
$$\eqalignno{
\{W_2,W_2\}_2 & = 2 W_2 \d' + W_2 \d \quad,\quad \{W_2,W_3\}_1 = 3 W_3 \d'
+ 2W_3' \d \quad,\cr
\{W_2,W_4\}_2 & = \frac23 W_2 \d''' + \frac43 W_2' \d'' + ( 4 W_4 + \frac23
W_2'' ) \d' + 3 W_4' \d \quad,\cr
\{W_3,W_3\}_2 & = \ha W_2 \d''' + \frac34 W_2' \d'' + ( 4 W_4 - \frac1{12}
W_2'' - 2 W_2{}^2) \d' + (2 W_4 - \frac16 W_2'' - W_2{}^2)' \d \quad,\cr
\{W_4,W_4\}_2 & = \ldots + ( 6 W_6 + \frac13 W_4'' + \frac1{45} W_2{}^{(4)}
- 6 W_4 W_2 - 6 W_3{}^2 - 2 W_2{}^3  - \ha {W_2'}^2 ) \d' \cr
&\qquad + ( 3 W_6 - W_4'' + \frac1{15} W_2{}^{(4)} - 3
W_4 W_2 - 3 W_3{}^2 - W_2{}^3  \cr
&\qquad \qquad + \ha (W_2^2)'' - \frac14 {W_2'}^2 ) \d
& \lee \cr }  $$

\noindent{\it Towards a representation of $\hat W_\infty$}

\noindent The algebra
$W_\infty$ was found to possess a free field representation in terms of two
real bosons \YannisII, so in \WYnineteen\ Yu and Wu  proposed
the existence of a similar representation of $\hat W_\infty$.
To this end, they introduced the currents $\j(x)$ and $\bj(x)$
with the natural Poisson brackets
\eqn\lf{
\{\j(x),\j(x')\}_2 = 0 ~,~ \{\j(x),\bj(x')\}_2 = \d'(x,x') ~,~
\{\bj(x),\bj(x')\}_2 = 0~.}
Upon imposing
\eqn\lg
{L = \pa + \sum_{i=0}^{\infty} u_i \pa^{-i-1} = \pa + \bj {1 \over \pa - (\j+
\bj)} \j\quad,}
we get an expression for the fields $u_i$ in terms of the currents, as
\eqn\lh{
u_i = \bj [(-\pa + \j + \bj)^i \ \j]\quad.}
If \lh\ were to provide a \it faithful \rm representation of $\hat
W_\infty$, one could just compute the $\{u_i,u_j\}$ commutators using their
expression in  \lh\ and the Poisson brackets \lf. However, more careful
examination \FFMR\ reveals that this is not the case. It is easy to check
that \lh\ implies relations or constraints between the fields $u_i$; the
simplest such relation is
\eqn\li{
u_2 u_0 - u_1^2 + u_1' u_0 - u_1 u_0' = 0 \quad. }
Thus it is not possible to unambiguously translate an expression in
$j~\bar{j}$ into an expression in the fields $u_i$. The upshot of this
is that the two boson representation is related to a {\it reduction},
as opposed to a {\it realization} of the KP hierarchy.

Let us show this more explicitly. The map between the fields $u_i$ and
$W_i$ was derived in \KP\
and gives, for the $\hat W_\infty$ case
\eqnn\lj
$$\eqalignno{
W_2 & = u_0\quad,\quad W_3 = u_1 + \ha u_0'\quad,\quad W_4 = u_2+u_1'
+ \frac13 u_0'' + u_0^2\quad,\cr
W_5 & = u_3 + \frac 32 u_2' + u_1'' + \frac14 u_0''' + 3 u_0 u_1 + \frac 32
u_0 u_0'\quad.\ & \lj \cr} $$
(note the correction of a typographical error
in the expression for $W_4$ in \KP).
In terms of $\j$ and $\bj$, we find
\eqnn\ljj
$$\eqalignno{
W_2 & = \j \bj \quad,\quad W_3 = \ha (\j \bj' - \j' \bj) + \j \bj^2 + \j^2
\bj \quad,\quad \cr
W_4 & = \frac13 (\j \bj'' - \j' \bj' + \j'' \bj) - \j \j' \bj +
\j \bj \bj' - \j' \bj^2 + \j^2 \bj' + \j^3 \bj + 3 \j^2 \bj^2 + \j \bj^3
\quad,\quad\cr
W_5 & = \frac14 (\j \bj''' - \j' \bj'' + \j'' \bj' - \j''' \bj)
+ \ha ( 2 \j^2 \bj'' + 2 \j'' \bj^2 + 3 \j \j'' \bj + 3 \j \bj \bj'' \cr
& \qquad + \j'^2 \bj + \j \bj'^2 - \j \j' \bj' - \j' \bj \bj' +
3 \j^3 \bj' - 3 \j' \bj^3 - 3 \j^2 \j' \bj + 3 \j \bj^2 \bj' \cr
& \phantom{\ha} \qquad - 9 \j \j' \bj^2 + 9 \j^2 \bj \bj' ) + \j^4 \bj +
6 \j^3 \bj^2 + 6 \j^2 \bj^3 + \j \bj^4
\quad,\ & \ljj \cr} $$
so that under the interchange $\j \rightarrow - \bj$, $\bj \rightarrow - \j$,
we see that $W_n \rightarrow (-)^n W_n.$ Such a symmetry was
already present for the bosonic representation of the \it linear \rm
$W_\infty,$ \YannisII. We note that the linear part (in terms of $u_i $
fields) of the field redefinitions \lj\ are given by a formula analogous
to \lh, namely
\eqnn\ljk
$$\eqalignno{
W_{n+2}^{\ lin}& = \{ u_0, u_1 + \ha u_0', u_2 + u_1' + \frac13 u_0'',
u_3 + \frac32 u_2' + u_1'' + \frac14 u_0''', \ldots \}
\cr
& = {1 \over n+1} \sum_{m=0}^n [ ( - \del + \j + \bj )^m \j]
[ ( \del + \j + \bj)^{n-m} \bj]  \quad.\ & \ljk\cr} $$
Given the relations \lj, \li\ becomes a constraint on the $W$ fields,
\eqn\lk{
W_4 W_2 = W_2^3 + W_3^2 - \frac14 W_2' W_2' + \frac13 W_2 W_2'' \quad.}
Note that the $\{W_4,W_4\}$ commutator in \lee\ involves a $W_4 W_2$
term, just like \lk.

\newsec{The $\j$, $\bj$ hierarchy}

In \KP\ a reduction similar to the one given in the last section was
considered at the level of the KP flows themselves. The second Hamiltonian
structure of this $\j$, $\bj$ hierarchy is clearly given by \lf, so we can
immediately write down the flows as
\eqn\ll
{\vec\j_{,t_r} = \pmatrix{\j \cr \bj \cr} _{t_r} = P_2 \nabla_{\vec\j} \int
\CH_r\quad,}
where $\nabla_{\vec\j} = (\d/\d\j,\d/\d\bj)$ and $P_2$ is the Hamiltonian
structure corresponding to \lf,
\eqn\lm
{P_2 = \bordermatrix{&\j&\bj\cr \j&0&\pa\cr \bj&\pa&0\cr}\quad,}
and $\CH_r$ is obtained by taking the expression for the KP Hamiltonian \lef\
and writing it in terms of the $\j$, $\bj$ fields through \lh.

Let us consider the second flow in more detail. It is written explicitly as
\eqnn\ln$$
\eqalignno{r = 2\quad:\CH_2 & = - \j' \bj + \j^2 \bj + \j \bj^2 \quad,\cr
\j_{,t_2} & = ( - \j'  + \j^2  + 2 \j \bj)' \quad, \cr
\bj_{,t_2} & = (\phantom{-}  \bj' + \bj^2 + 2 \j \bj)' \quad.\qquad \qquad
\qquad & \ln\cr}$$
The $\j$, $\bj$ hierarchy turns out to be bi-Hamiltonian, with the first
structure $P_1$
(corresponding to $W_{1+\infty}$)  being non local, but the third
Hamiltonian structure $P_3 = P_2 P_1{}^{-1} P_2$ being local.
$P_3$ and $P_1$ are given explicitly in \KP.

Let us try to manipulate \ln\ to see if it can be made to correspond to
a known integrable system. We notice that upon setting $\bj$ to $0$, \ln\
becomes the ``derivative'' of the Burgers equation, $h_t = h'' + 2 h h',$
which can be linearized by the Cole--Hopf transformation $h = u'/u$ into
the heat equation
$u_t = u''$ \Olver. Guided by this analogy, let us first ``integrate''
the flows \ln\ by introducing $h$ and $\bar h$ defined by
$h' = - \j$ and $\bar h'=\bj.$ We get:
\eqnn\lop
$$
\eqalignno{
     h_{,t_2} & = - h'' - h'^2 + 2 h' \bar h' \quad, \cr
\bar h_{,t_2} & = \phantom{-} \bar h''  + \bar h'^2 - 2 h' \bar h'
\quad.\qquad
& \lop\cr}$$
Using these equations, we find that $\psi = h' e^{h-\bar h} $ and
$\bar\psi = \bar h' e^{\bar h- h}$ satisfy
\eqnn\lo$$
\eqalignno{
\psi_{,t_2} & = - \psi'' + 2 \psi^2 \bar\psi \quad, \cr
\bar\psi_{,t_2} & = \phantom{-} \bar \psi'' - 2 \bar \psi^2 \psi
\quad,\qquad & \lo\cr}$$
which is the second flow of the NLS system. A more careful treatment of the
relation between our $\j$, $\bj$ system and the NLS one in standard form can
be found in \JerNLS. In fact it can be checked that the entire
$\j$, $\bj$ hierarchy can be mapped in this manner to the
NLS hierarchy. The simplest way to do this is to exploit the powerful
concept of Fr\'echet derivatives (see for instance \Olver) to
map the different Hamiltonian structures of the $\j$, $\bj$ hierarchy to
those of the NLS hierarchy. From $h = - \del^{-1} \j$ and
$\bar h = \del^{-1} \bj,$
we find that the Fr\'echet derivative of  $(h, \bar h)^T$ with respect to
$(\j,\bj)^T$ is
\eqn\loo
{D = \pmatrix{ - \del^{-1} & 0 \cr 0 & \del^{-1} \cr } }
so that the Hamiltonian structure of the flows \lop\ is
\eqn\loq
{D P_2 D^{\dagger} =
\pmatrix{ - \del^{-1} & 0 \cr 0 & \del^{-1} \cr }
\pmatrix{ 0 & \del \cr \del & 0 \cr}
\pmatrix{ \del^{-1} & 0 \cr 0 & - \del^{-1} \cr}  =
\pmatrix{ 0 &\del^{-1} \cr \del^{-1} & 0 \cr} }
(Here we are assuming the possibility of defining $\del^{-1}$ as an
anti-self-adjoint operator). The Fr\'echet derivative of
$(\psi,\bar\psi)^T$ with respect to $(h,\bar h)^T$
is
\eqn\lor
{\tilde D = \pmatrix{ {e^{h-\bar h}(\del + h')} &
                -{e^{h-\bar h} h' }          \cr
                -{e^{\bar h -h}\bar h' } &
                 {e^{\bar h -h}(\del + \bar h')} \cr }  }
and using this we find
\eqn\los
{\tilde D D P_2 D^{\dagger} \tilde D^{\dagger}
= \pmatrix{ - 2 \psi \del^{-1} \psi & - \del
+ 2 \psi \del^{-1} \bar \psi \cr
- \del + 2 \bar\psi \del^{-1} \psi &  - 2 \bar \psi \del^{-1} \bar \psi \cr }
}
This is the second Hamiltonian structure of the NLS system. Similarly
one can show explcitly that the $P_1$ induces the first
Hamiltonian structure of NLS.

We can formulate the connection between the KP and NLS hierarchies
directly, without going through the $j, \bar{j}$ hierarchy.
It is straightforward to show that the Lax operator (1.9) can be
written
\eqn\lq
{L = \del + \bj (\del - \j - \bj)^{-1} \j
  = \del - \psi \del^{-1} \bar\psi  }
by simply using the map from the $\j$'s to the $\psi$'s we just presented.
So the reduction from KP to NLS is given by the constraint
$L_{KP} = \del - \psi \del^{-1} \bar\psi$, or, equivalently,
by constraining the KP fields as $u_i =
(-)^{i+1} \psi \bar\psi^{(i)}.$ A natural question is to ask what kind
of hierarchies one gets by considering
the reduction $L_{KP}^n = \del^n + v \del^{n-2} + \ldots + \psi \del^{-1}
\bar\psi.$ This has been considered in some details in \OS.

Finally the relation between the NLS hierarchy and the $\j ,\bj$ system
allows us to answer the question that was left open in \KP, of finding a
zero-curvature formulation of the
$\j$, $\bj$ hierarchy, and of how to obtain it  by reduction from
the self--dual Yang--Mills system \MSBDNS.
Since we use slightly different notation than in \KP, we recall how
such a reduction is introduced.
The self--dual Yang--Mills equations in four dimensions are usually
written as
\eqn\lqq{F_{12} = F_{34} \quad,\quad F_{13} = F_{42} \quad,\quad
F_{14} = F_{23} \quad,}
where $F_{\mu\nu} = \del_\mu A_\nu - \del_\nu A_\mu + [ A_\mu,A_\nu ].$
Introducing $w = x^1 + i x^2,$ $z = x^3 - i x^4$ and their complex conjugates
$\bar w$ and $\bar z,$ eqs.\lqq\ become
\eqnn\lr$$
\eqalignno{
F_{\bar z \bar w} & = 0 \quad,\qquad\cr
F_{z \bar z} + F_{w \bar w} & = 0 \quad,\qquad\cr
F_{ z w} & = 0 \quad,\qquad
\qquad & \lr\cr}$$
where now $F_{zw} = \del_z A_w - \del_w A_z + [ A_z,A_w ].$ The gauge
freedom in \lr\ is expressed through the invariance of \lr\ under the
transformations
$A_\mu \rightarrow A_\mu' = g A_\mu g^{-1} - \del_\mu g \, g^{-1}.$
Now, we know that to get the NLS equations by reduction of the self--dual
Yang--Mills system, we just need to choose
\eqn\lrr{A_z = \pmatrix{ 0 & \psi \cr \bar\psi & 0 \cr} \quad,\quad
A_{\bar w} = \ha \pmatrix{ 1 & 0 \cr 0 & -1 \cr} \quad.}
and reduce the system \lr\ first with respect to $\del_{\bar w} $ and
then $\del_z - \del_{\bar z}$
\MSBDNS. The potentials $A_{\bar z}$ and $A_w$
are determined by \lr. Here, since we know that the $\j ,\bj$ system is
related to the $\psi ,\bar\psi$ system by the map given above, we consider
the effect of a $\bar w$-independent
gauge transformation on $A_z$ which leaves $A_{\bar w}$
invariant. Under a transformation by
\eqn\lrs{g = \pmatrix{\a^{-1} & 0 \cr 0 & \a \cr } \quad,}
we get
\eqn\lrt{A_z' = \pmatrix{\a^{-1} \a_z & \psi \a^{-2} \cr \bar\psi \a^2 & -
\a^{-1} \a_z\cr } = \pmatrix{ - \ha ( \j + \bj) & - \j \cr \bj &
\ha(\j + \bj) \cr} \quad, }
where we have taken $\a^2 = e^{\bar h - h}$.

If we take the new form of $A_z$ from \lrt,
and the form of $A_{\bar w}$ from \lrr\ and plug into
\lr, imposing $\del_{\bar w}=0$ but {\it not} $\del_{z}=\del_{\bar{z}}$,
we find
\eqnn\lru$$
\eqalignno{
A_{\bar z} & = \b \pmatrix{ 1 & 0 \cr 0 & 1\cr}\qquad{\rm with\ }
\b_z = -\ha (\j + \bj)_{\bar z}\cr
\pmatrix{ \j \cr \bj \cr}_z & = \pmatrix{ - \del + \j_z \del^{-1}
+ (\j + \bj) & \j_z \del^{-1} + 2 \j \cr
\bj_z \del^{-1} + 2 \bj & \del + \bj_z \del^{-1} + (\j + \bj) \cr}
\pmatrix{\j \cr \bj \cr}_{\bar z}
\qquad.\quad
& \lru\cr}$$
The matrix operator  in the second equation above
is equal to $P_3 P_2^{-1}$ in \KP. The method
of reduction used here, not imposing immediatly $\del_{z}=\del_{\bar{z}}$,
gives us not only the equations of the hierarchy we are looking
for, but also its recursion operator (of course, the $r=2$ flow \ln\ is
obtained after setting $\del_z = \del_{\bar z}$). The same applies to
the reduction to the KdV equation, where one gets not only the usual
KdV equation, but also its recursion operator.

\newsec{Some open questions}

The results presented here open up some potentially interesting avenues of
research. The $\hat W_\infty$ algebra, which was originally introduced
as an algebra that ``contains'' all the $W_N$ algebras, apparently
contains even more. It is clearly of some interest to try to find all
reductions of the KP system and to understand the physical significance
of the associated reductions of $\hat W_\infty$. Work on the algebras
associated with the class of reductions of \OS, generalizing the
result here, is currently in progress.

Another question that arises
is as follows: we know that the KdV equation arises
by reduction (i.e. constraining) the KP hierarchy. On the other hand,
we also know how to reduce the self-dual Yang-Mills system to the KdV
equation \MSBDNS.
In the work presented here, we see that the KP hierarchy can also be reduced
to the NLS hierarchy, and again we know how to reduce
the self--dual Yang--Mills equations to NLS. It is believed that self--dual
Yang--Mills is a universal integrable system, and in particular KP can
be obtained by reduction from it, for a suitable infinite dimensional
gauge group. It would be interesting to understand better why certain
integrable systems can appear both as direct reductions of self--dual
Yang--Mills with a finite dimensional gauge group, and as indirect
reductions via KP. Work on this is also in progress.

\vskip2truecm
{\bf Acknowledgements}\par

We wish to thank  Pierre Mathieu and Walter Oevel for discussions.

D.A.D was supported by NSERC (Canada), FCAR
(Qu\'ebec), and BSR (Universit\'e Laval), and J.S. was
supported by a grant in aid from the U.S. Department of Energy, \#
DE-FG02-90ER40542.

\listrefs

\end